\documentclass[prl,twocolumn,showpacs,superscriptaddress]{revtex4}

\usepackage{graphicx}
\usepackage{color}

\begin{document}

\title{Combining high pressure and coherent diffraction: a first feasibility test}

\author{D. Le Bolloc'h}
\affiliation{Laboratoire de Physique des Solides (CNRS-UMR 8502), B{\^a}t. 510, Universit{\'e} Paris-sud, 91405 Orsay cedex, France}
\author{J.P. Iti\'e}
\affiliation{Synchrotron SOLEIL - L'Orme des Merisiers Saint-Aubin - BP 48 91192 GIF-sur-YVETTE, France}
\author{A. Polian}
\affiliation{IMPMC, Campus Boucicaut 140, rue de Lourmel, 75015 Paris, France}
\author{S. Ravy}
\affiliation{Synchrotron SOLEIL - L'Orme des Merisiers Saint-Aubin - BP 48 91192 GIF-sur-YVETTE, France}

\begin{abstract}
We present a first experiment combining high pression and coherent X-ray diffraction.
By using a dedicated diamond anvil cell, we show that
the degree of coherence of the X-ray beam is preserved when the X-ray beam passes through the diamond cell. 
This observation opens the possibility of studying the dynamics of slow fluctuations under high pressure.
\end{abstract}

\maketitle

\section{Introduction}

Coherent diffraction techniques provide a new insight into a wide variety of phenomenon. 
In particular, disordered samples which are coherently illuminated give rise to speckle pattern 
and speckle fluctuations analysis gives access to the dynamic of fluctuations of matter. 

{\it In situ} experiments using various external constraints as temperature, external current or magnetic field
can be easily combined with coherent x-rays as far as the sample environment does not destroy 
the coherence properties of the beam. Up to now this has not been demonstrated for diamond anvil cell. In classical anvil cell, the x-ray beam goes through the diamonds or the gasket which may induce an irreparable loss of coherence.

We show in this paper that high pression and coherent diffraction can both be combined, going through the diamond anvils. We exclude the other geometry, through a Be gasket for example, because such materials with non perfect surfaces strongly affect the degree of coherence of the beam. This is not the case of the diamond anvils which are nearly perfect single crystals with a well polished surface. In that case, we show that
the degree of coherence of the beam is almost not perturbated when the beam passes through the diamond cell.
This setup opens the possibility to study time fluctuations under high pressure.

\section{Propagation properties of the transverse coherence length}

Thanks to third generation synchrotron radiation sources, it is possible to obtain X-ray beams with
an excellent degree of coherence. This experimental fact may be surprising since synchrotron sources are 
only weakly coherent. In fact, this is possible thanks to the propagation properties of the transverse coherence length.
The propagation properties of a partially coherent beam can be obtained from the mutual coherence function
$$\gamma(\vec r_1,\vec r_2,z_0)=\left\langle A(\vec r_1)A^*(\vec r_2)\right\rangle,$$
which characterizes the spatial coherence properties of the amplitude $A(\vec r)$ of the field at a pair of points $(\vec r_1,\vec r_2)$,
 located in the plane
perpendicular to the direction of propagation. If
the beam profile at $z_0$ 
is assumed to be Gaussian, in the Schell approximation\cite{born}, 
an analytical expression of the transverse coherence length $\xi(z)$ can be obtained:
  \begin{eqnarray}
  \label{xiz}
  \xi^2(z)=\xi_0^2+\left(\frac{z\lambda}{2\pi}\right)^2 \left(\frac{\xi_0^2+2\sigma^2}{\sigma^4}\right).\nonumber
  \end{eqnarray}
This last equation gives the transverse coherence length $\xi(z)$ of the beam at any distance $z$ from the source 
as a function of the beam size $\sigma$ and of the coherence properties of the source $\xi_0$.
For example, in the case of a totally incoherent source $\xi_0=0$, the transverse coherence length 
$\xi(z)$ will not be zero:
  \begin{eqnarray}
  \xi(z)\propto \frac{z\lambda}{\sigma}.
  \end{eqnarray}
As an example, if the source is totally incoherent and
located at 40m from the sample position, we should obtain
transverse coherence lengths up to $\xi_v(z)\approx 200\mu m$ (with $\lambda=1.5\AA$ and $\sigma_{rms}=10 \mu m$).
Unfortunatly, optical aberrations, mainly due to focussing mirrors, reduce considerably this value.
In practice, secondary slits are located $10 m$ upstream from the sample positon with 100-250 $\mu m$ apertures 
in order to clean the wavefront. We then obtained $\xi_v(z)$ around $10\mu m$ at the sample position.
We just need to reduce the beam size close to $\Phi=10\mu m$ to obtain a degree of coherence $\beta=\xi_v/\sigma$ 
which tends towards 1.
\begin{figure}
\begin{center}
\resizebox*{9cm}{!}{\includegraphics{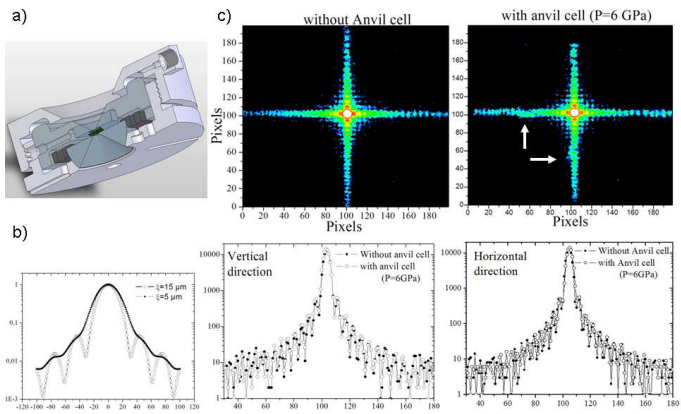}}
\caption{a) Design of the diamond anvil cell. b) Theoretical cross-like diffraction profile of a rectangular slit for two values of the degree of coherence $\xi_v$ (a=$10\mu m$). c) Contrast of fringes of the cross-like diffraction of a rectangular slit ($\Phi=5\mu m$), 
with and without the anvil cell along the optical path.
The two white arrows indicate two slight deformations of the profile of fringe induced by the anvil cell.}
\label{figure0}
\end{center}
\end{figure}

\section{Diffraction pattern of a squared slit from a pratially coherent X-ray beam}

The mutual coherence function
can be used to reproduce the diffraction pattern of a squared slit as a function of the 
coherence properties of the x-ray beam.
In the case of a totally coherent beam, we obtain  the well-known cross-like intensity profile:
\begin{equation}
I(x)=I_0\frac{\sin^2{(\frac{\pi a}{z\lambda}x)}}{(\frac{\pi a}{z\lambda}x)^2},
\label{coherence}
\end{equation}
where $a$ is the slit aperture.
As $\xi$ decreases, the contrast of fringes is getting more and more reduced as displayed in figure 1a).
Provided that the slit blades are carefully polished\cite{lebolloch}, the measurement of the 
contrast of fringes gives a good estimation of the degree of coherence.
In the following, we have measured how a coherent beam propagates through the anvil diamond cell 
from the contrast of fringes of a rectangular slits diffraction pattern. 

 \subsection{Combining pressure and coherent diffraction}

To combine high pressure and coherent diffraction will provide an unique way to measure 
fluctuations under high pressure. From an experimental point of view, the combination of both techniques is not 
obvious. A good compromise has to be found between the degree of coherence (which decreases with increasing photons energy), the absorption of the anvils, the  opening of the cell and the maximum pressure expected. The geometry of the diamond anvil cell is displayed in figure 1b). We use 1.2 mm thick diamonds and 10 keV x-rays. Compared to 8 keV x-rays usually used in coherent diffraction experiments, the degree of coherence is reduced by $20\%$ (see Eq.1)  but the transmission is increased by a factor 3. The cell is designed to have an opening of $2\times40^{\circ}$ with such diamonds. 
 Diffraction patterns of rectangular slit has been recorded with and without the presence of the anvil cell at the sample position (see figure 1c).
 The secondary slits were closed at $100\mu m$ and the slit located at the sample position at $\phi=5\mu m$. 
 The 2D images have been recorded with a CCD camera ($13\mu m$ pixels size) located at 2m from the sample position.
 The degree of coherence without diamond cell has been estimated from the contrast of fringes around $\beta=70\%$.
 \begin{figure}
\begin{center}
\resizebox*{9cm}{!}{\includegraphics{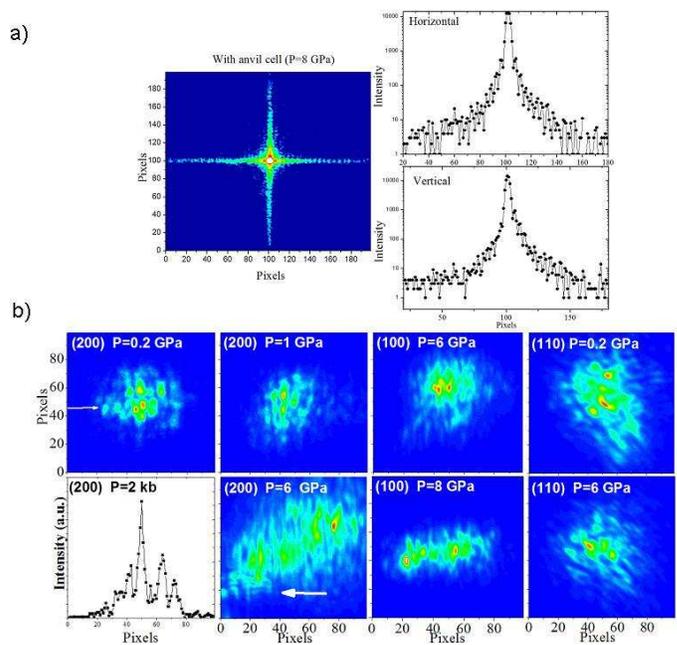}}
\caption{a)For larger pressure (P=8GPa), the cross-like intensity profile is more disturbed ($\Phi=6.5\mu m$). b) Diffraction patterns of the (200), (100) and (110) Bragg reflections in $SrTiO_3$ at various pressure up to 8GPa.
The speckle patterns were obtained with F5=$100 \mu m\times100\mu m$ and $\Phi=10\mu m\times 10\mu m$.
The profile of fringes of the (200) Bragg peak at P=0.2GPa is also displayed. At P=6 Gpa, the (200) bragg reflection is
cut by the anvil cell aperture (see arrow) at $2\theta=38^o$.}
\label{figure1}
\end{center}
\end{figure}
 The main result of this paper is to establish that when using such anvil cell, with a residual pressure of 0.2GPa, similar contrast of fringes is observed.
The coherence properties of the beam are not altered by our anvil cell. This opens the possibility to do coherent diffraction experiment under high pressure.
 In fact, the contrast of fringes is slightly stronger with the anvil cell (see figure 1c). This is probably due to the fact that the anvil cell
 absorbs X-rays scattered by air or kapton foils upstream. Note also that the diffraction pattern is locally perturbed
 at two positions along the cross-like profile (see arrows in figure 1c).
 
A similar experiment has been performed at higher pressure, up to 8GPa.
Unfortunatly, the figure 2a has not been exactly obtained in the same configuration. The secondary slits were opened at $200\mu m$ and the last slits at $\Phi=6.5\mu m$. The contrast of fringes is thus smaller. Nevertheless, it seems that for
larger pressure, the contrast decreases.
We can not exclude that the degree of coherence decreases for higher pressure due to larger deformations of the diamond cell. 

As an example of coherent diffraction at wide angles, we choose the $SrTiO_3$ system which has been recently study by this method close to
its displacive transition\cite{ravy}. We have measured different fundamental Bragg peaks, in the scattering plane (the (100) and (200) reflections) and out-of-plane (the (110) reflection), at several pressures (see figure 2b).
Reflections up to the (200) are reachable thanks to the large angular aperture of our diamond cell. 
We obtained speckles with a good contrast up to 8GPa.
The micrometer-size $SrTiO_3$ sample was not a perfect monocrystal. Speckles are probably due to interferences between misoriented domains. In addition, the crystal broke up into several pieces under higher pressure.

\medskip

\end{document}